\documentclass[12pt]{article}
\usepackage{pdproc,epsfig} 

\begin{document}

\def\permille{$^\circ/_{\circ \circ} $}
\def\lapprox{\mathrel{\mathop  {\hbox{\lower0.5ex\hbox{$\sim$}
\kern-0.8em\lower-0.7ex\hbox{$<$}}}}}
\def\gapprox{\mathrel{\mathop  {\hbox{\lower0.5ex\hbox{$\sim$}
\kern-0.8em\lower-0.7ex\hbox{$>$}}}}}
\renewcommand{\thefootnote}{\alph{footnote}}
\def\antinu{\bar{\nu}}
  
\title{NEUTRINOS FROM SAN MARCO AND BELOW}

\author{G. FIORENTINI}

\address{Dipartimento di Fisica, Universit\`a di Ferrara and INFN-Ferrara,
  Via Paradiso 12\\
 Ferrara, I-44100 Italy\\
 {\rm E-mail: fiorentini@fe.infn.it}}

\author{M. LISSIA}

\address{Dipartimento di Fisica, Universit\`a di Cagliari and INFN-Cagliari,
 Strada provinciale per Sestu Km 0.700  \\
 Monserrato (Cagliari), I-09042 Italy\\ }

\author{F. MANTOVANI}

\address{Dipartimento di  Scienze della Terra, Universit\`a di Siena,
 Via Laterina 8\\
 Siena, I-53100 Italy\\ 
 and INFN-Ferrara, Via Paradiso 12\\
 Ferrara, I-44100 Italy}

 \centerline{\footnotesize and}

\author{B. RICCI}

\address{Dipartimento di Fisica, Universit\`a di Ferrara and INFN-Ferrara,
  Via Paradiso 12\\
 Ferrara, I-44100 Italy}



\abstract{
Order of magnitude estimates  of radiogenic heat  and  
antineutrino production are given, using the San Marco cathedral 
as an example. Prospects of determining  the radiogenic contribution 
to terrestrial  heat by detection of antineutrinos from natural 
radioactivity (geoneutrinos) are discussed.  A three kton scintillator 
detector in three years can clearly discriminate among different models 
of terrestrial heat production.  
In addition, the study  of geoneutrinos offers a possibility 
of improving the determination 
of  neutrino mass and mixing, by exploiting the knowledge of 
Th/U abundance in the Earth.
}

\normalsize\baselineskip=15pt

\section{A few facts about San Marco}

After the  visit of the San Marco cathedral, accompanied by a most 
interesting historical and artistic overview, nothing should be added but our 
gratitude to Milla Baldo Ceolin for organizing this most interesting and 
timely conference and for spicing it with exceptional events. Physicists 
however are tenacious, and we cannot  resist adding a few additional 
information, some of these a touristic guide will never tell you.

First of all, San Marco contains radioactive materials. The cathedral 
mass being in the range of 100 kton , we expect it contains about 
100 kg of Uranium, as the typical Uranium abundance in  rocks is 
in the range of one part per million, however with large variations.  
We  also expect some  400 kg of Thorium, since  almost everywhere 
in the solar system (meteorites, Moon, Venus and also Earth)  
the typical abundance ratio is $ Th/U \simeq 4$. In addition, 
there are about 100 Kg of 
$^{40}K$,  corresponding to the typical 
ratio  $K/U \simeq 10^4$ which one finds in Earth rocks and to 
the natural abundance $^{40}K/K = 1.2 \cdot 10^{-4}$.

San Marco is also a  heat source.  In fact, each  decay chain 
releases energy over long time scales  ({\em e.g.} for $^{238}U$  
$\Delta= 52$ MeV  and  $\tau_{1/2}=4.5$ Gyr ), the total heat flow being: 
\begin{equation}
\label{heat}
H	= 9.5 M(U) + 2.7  M(Th) + 3.6 M(^{40}K)
\end{equation}
with masses  in $100$ kg and $H$ in mW. This gives 24 mW for San Marco,  
really a very weak heat source, although it does not matter in 
these sunny days.

More  interesting to people attending this conference, 
San Marco is an anti-neutrino source. Each  decay chain 
releases antineutrinos  together with heat, with a well fixed ratio. 
({\em e.g.} $^{238}U \rightarrow ^{206}Pb+8\, ^4He +6 \antinu +52$ MeV). 
The antineutrino luminosity is:
\begin{equation}
\label{lum}
L	= 7.4 M(U) + 1.6  M(Th) + 27 M(^{40}K)
\end{equation}
where again $M$ is  in 100 Kg and $L$ in $10^9 \antinu /s$. 
This gives about $4\cdot 10^{10} \antinu /s$ for San Marco.
Antineutrinos from the progenies of Uranium  ($E_{max}=3.3 MeV$) 
and Thorium ($E_{max}=2.2 MeV$)   can be detected by means of 
inverse beta decay reaction:
\begin{equation}
\label{inverse}
\antinu + p \rightarrow n+ e^+ -1.804\, MeV
\end{equation}

At least in principle, the two components can be discriminated, 
due to the different end points.  Unfortunately, antineutrinos 
from $\beta$ decay of $^{40}K$ are below the threshold for (\ref{inverse}), 
whereas neutrinos from electron capture are obscured by the Sun.
The cross section of (\ref{inverse}) for 2.5 MeV antineutrinos 
( $\sigma \simeq 5 \cdot 10^{-44}cm^2$) corresponds  to an interaction 
length in water $\lambda= 7 \cdot 10^{18}$ m.  There is plenty of water 
near the cathedral and San Marco  square ($S \simeq 10^4$ m$^2$) 
is often covered with Acqua Alta (high water),  a 10 cm  height 
corresponding  roughly to 1 kton, the size of KamLAND. 
Should Acqua Alta reach the clock of Torre dell'Orologio,  the size 
of Superkamiokande is reached.  With a height of 100m  a megaton detector 
is obtained, the cathedral being now  deeply  submerged. 
A pointlike source emitting $10^{10}\antinu /s$ with E=2.5 MeV  
at the center of a megaton water sphere will produce one event every 
four years. Also in view of the  environmental impact of such a project, 
better we look  at some other direction.

\section{ The sources of  terrestrial heat}

Earth re-emits in space the radiation coming from the sun 
($K_\odot =1.4$ kW/m$^2$)  adding to  it a a tiny flux of heat produced  
from its   interior ($\Phi \simeq 80 mW/m^2$). By integrating 
this latter over the Earth surface one gets a flow $H_E = 40TW$, 
the  equivalent of  some $10^4$ nuclear power plants.

The origins of terrestrial heat are not understood in quantitative 
terms.  In 1980  J. Verhoogen concluded a  review of terrestrial 
heat sources by saying \cite{Ver}:
{\em ``...what emerges from this morass of fragmentary and uncertain 
data is that radioactivity itself could possibly account for at 
least 60 per cent if not 100 per cent of the Earth's heat output.
If one adds  the greater rate of radiogenic heat production in the 
past, possible release of gravitational energy 
(original heat, separation of the core...) 
tidal friction ... and possible meteoritic impact ... 
the total supply of energy may seem embarassingly large..."}.
One can appreciate the complexity of the problem by comparing Sun and 
Earth energy inventories. In fact,  a constant  heat flow $H$ can be 
sustained by an energy source $U$ for an  age  $t$ provided that 
$U\ge H t$.  For sustaining  the sun over an age 
$t_{solar} =4.5 \cdot 10^9$ yr, gravitational and chemical energies are 
short by a factor $10^2$ and $10^6$ respectively  and only nuclear 
energy can succeed, as beautifully demonstrated by Gallium experiments
 in the nineties. 
On the other hand,  the terrestrial heat can be sustained over 
geological times by any  energy source, be it nuclear, gravitational 
or chemical. 

Observational data on the amounts of Uranium, 
Thorium and Potassium in Earth interior  are rather limited, 
since only the crust and the upper part of 
the mantle are accessible to geochemical analysis.  
As $U, \, Th$ and $K$
 are lithofile elements, they accumulate in the  continental crust (CC).  
Estimates for the Uranium mass in the crust are in the 
range\cite{Taylor,Wed}
\begin{equation}
\label{Mcrust}
  M_c(U)= (0.2-0.4)10^{17}\,kg\quad .
\end{equation}  

Concentrations in the mantle are much smaller, however the total 
amounts are  comparable due to the much larger mass of the mantle.  
Estimates for the whole mantle are in the range\cite{Ver}
\begin{equation}
\label{Mmantle}
 M_m(U)= (0.4-0.8)10^{17}\,kg \quad .
\end{equation}

One has to remark, however, that these estimates are much more uncertain 
than for the crust 
as  they are obtained by: i)collecting data for upper mantle ($h_u=600km$),
ii)extrapolating them to the completely  unexplored  lower mantle 
($h_l=3000km$). 

Concerning the abundance ratios, one has generally  $Th/U \simeq 4$, 
consistent with the meteoritic value. A remarkable exception is the 
oceanic crust  where $Th/U \simeq 2$,   however  both $U$ and $Th$
 abundances are an order of magnitude smaller with respect  to CC, 
which is also much thicker. 

Concerning Potassium, generally one finds $K/U\simeq 10,000$. Earth 
looks thus  significantly  impoverished in Potassium with respect 
to Carbonaceous Chondrites,  and   also to other meteorites. 
This has long been known as the Potassium problem\cite{Brown1,Brown2}.
 In fact, 
elements as heavy as Potassium should not have escaped from a planet as
 big as Earth.
It has been suggested that at high pressure Potassium behaves as a 
metal and thus it could have been buried  in the Earth core, 
where it could provide the energy source of the terrestrial magnetic field.  
However, Potassium depletion is also observed in Moon and 
Venus  rocks. The most reasonable assumption is that 
it volatized in the formation of  planetesimals from which Earth 
has accreted. 

In conclusion, the determination of the Uranium, 
Thorium and Potassium in the Earth is an important 
scientific problem, as it can fix the radiogenic contribution 
to terrestrial heat production.

\section{ Antineutrinos from below }
 
``{\em If there are more things in heaven and Earth than are dreamt 
of in our natural philosphy, it is partly because electromagnetic 
detection alone is inadequate}''.  With these words in 1984, 
Krauss, Glashow and Schramm\cite{Krauss}  proposed a 
program of antineutrino astronomy and geophysics, which could 
open vast new windows for exploration above us and below. Now 
that we understand the fate of neutrinos it is time to tackle 
the program (including  Earth energetics, a detailed study of the 
solar core,  neutrinos from past supernovae...) Determination of  
the radiogenic contribution to terrestrial heat production is the  
first step. 

One can build several models for the radiogenic heat production. 
Since  for any element there is a well fixed  ratio heat/(anti)neutrinos 
each model also provides a prediction for the antineutrino luminosities, 
the basic equations being (\ref{heat}) and (\ref{lum}), 
where the same numerical 
coefficients can be used when masses are in units of $10^{17}$ kg, powers 
are in TW and luminosities in  $10^{24}/s$. At this level, everything is 
fixed in terms of three inputs. The range of plausible models is 
covered  in  fig.\ref{fig1}, which deserves the following comments: 

\begin{figure}
\centering
         \mbox{\epsfig{figure=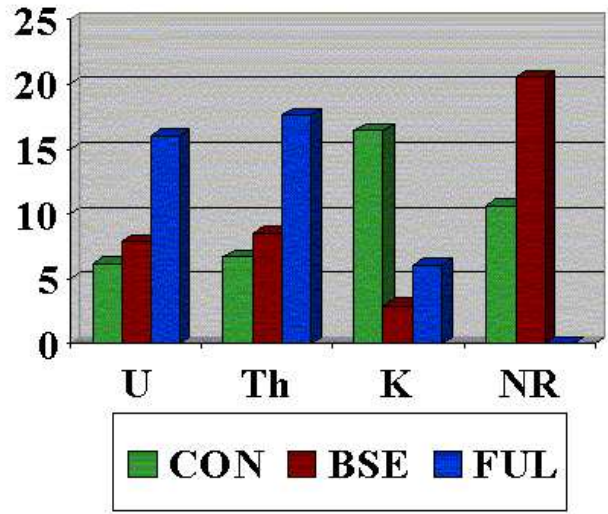,width=10.0cm}}
\caption{Contributions to terrestrial heat production (TW), 
as estimated in different models of Earth interior. 
CON= chondritic, BSE=bulk silicate Earth, RAD= fully radiogenic. 
The non radiogenic contribution is indicated as NR.}
\label{fig1}
\end{figure}

i)In  a  simple chondritic model  one assumes that Earth 
is obtained by assembling together the same material as 
we find in these meteorites, without loss of heavy enough elements. 
The amounts of $U$, $Th$ and $K$ are determined from their ratio to $Si$  in
 meteorites, by rescaling to the known abundance of this latter 
element in the Earth. This simple model easily  accounts for 3/4 of $H_E$ ,  
mainly supplied  from $^{40}K$, however it implies $K/U= 7\cdot 10^4$, 
a factor seven larger than the value observed in Earth rocks.

ii) In the standard model of geochemists, the so called  
Bulk Silicate Earth (BSE) model, with  $K/U= 1\cdot 10^4 $, 
the radiogenic production is  about one half of the terrestrial 
heat flow, being  supplied  mainly from U and Th.

iii) A fully radiogenic model, where the BSE abundances are 
rescaled imposing by imposing  $H_{rad}=40$ TW, is not excluded 
by observational data.

\begin{figure}
\centering
         \mbox{\epsfig{figure=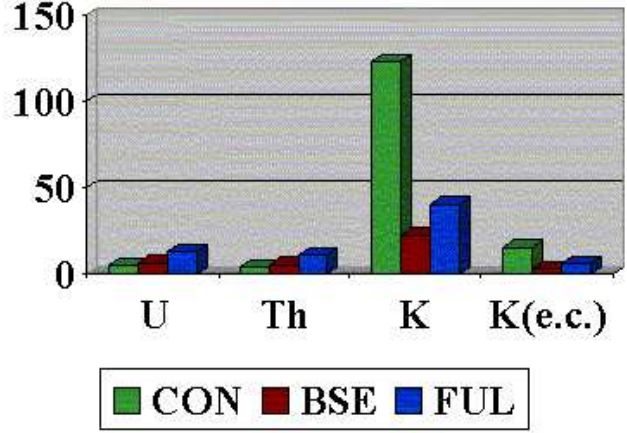,width=10.0cm}}
\caption{{\bf Neutrino luminosity}.
Antineutrinos production rates from U, Th and K according to 
different models of Earth interior. In the last column neutrinos 
from e.c. of $^{40}K$ are shown.  Units are $10^{24}$s$^{-1}$.}
\label{fig2}
\end{figure}

In all models, see fig.\ref{fig2}, antineutrino production is 
dominated by $^{40}K$ decays. Th and U anti-neutrino luminosities 
are in the range $(10-20)\cdot 10^{24}/s$. By dividing over the earth 
surface, one gets fluxes of order $10^6$ cm$^{-2}$s$^{-1}$, 
in the same range as that of solar boron neutrinos.

Clearly, in order to estimate fluxes at a specific site 
one needs assumptions about the distribution of the radioactive materials, 
the total amounts being an insufficient information.  In a paper submitted 
on December 2002 we provided estimates corresponding to different models 
for several sites\cite{Mantovani}. 
The numbers of predicted events for Kamland 
(normalized to an exposure of  $0.14\cdot  10^{32} p\cdot yr$,   
a detection efficiency  $\epsilon=78\% $ and  a survival probability 
$P_{ee}=0.55$)  are 3.5, 4 and 6 for the chondritic, 
BSE  and fully radiogenic model respectively.

The paper ended with: 
{ \em ``The determination of the radiogenic component of terrestrial heat 
is an important and so far unanswered question.... , the first fruit
 we can get from neutrinos, and Kamland will get the firstlings very soon"}.

A few days later the first results from Kamland appeared\cite{KamLAND},  
containing a first glimpse of Earth interior in addition 
to important information on neutrino oscillation. 
Out of a total of 32 counts,
in the prompt energy region below 2.6 MeV,  
20 events are 
associated with antineutrinos from reactors and 3 correspond to 
the estimated background. The remaining  9 counts are the first 
indication of geo-neutrinos! Clearly the uncertainty is large, since 
the expected statistical fluctuations are about $\sqrt{32}$. 
Within this error,  the result is consistent with any model for 
the radiogenic contribution to terrestrial heat, from 0 to 100 TW.   
One has to remind that data have been collected in just six months 
and that significant accumulation can be achieved in the future.

\section{Prospects for measuring terrestrial heat with geoneutrinos}

The first Kamland results really open a new window on Earth interior.  
A natural question is thus: what can be  learnt on heat production 
from geoneutrino measurements?

Let us consider  Uranium geoneutrinos, since  expected  event 
numbers are higher and  energies can be chosen in a range where  
background is negligible. The basic equations relating event number $N(U)$,  
Uranium mass in the crust $M_c(U)$, in the mantle $M_m(U)$,  and  
Uranium contribution to heat production rate $H(U)$ are:
\begin{equation}
\label{heat2}
H(U)=  h [ M_c(U)+ M_m (U)]
\end{equation}
\begin{equation}
\label{events}
N(U)= n_c M_c(U) + n_m M_m (U).
\end{equation}

The first equation simply adds the contribution of crust  and mantle, 
the proportionality constant being $h= 9.5TW/10^{17}$ kg. 
The different coefficients  $n_c$ and $n_m$  reflect the different distances 
of the detector from sources in the crust  and in the mantle. 
They  also depend on the detector characteristics and on neutrino properties:
as a normalization we shall consider an
exposure of  
$10^{32} p \cdot yr$\footnote{1kton of mineral oil contains
 $0.86 \cdot 10^{32}$ free protons.},
a  100\% detection efficiency and  a neutrino survival probability 
$P_{ee}=0.55$.
 
Assuming a uniform distribution in the mantle one has,  independently 
of the site, $n_m =18/(10^{17}$kg). On the other hand, $n_c$ depends on the 
location. 
Our preliminary calculations, based on a crustal map of the whole Earth, 
give $n_c=52/(10^{17}$kg) and  $n_c=63/(10^{17}$kg) for the Kamioka 
mine and the Gran Sasso respectively. 

There are two unknowns in eqs.(\ref{heat2}) and (\ref{events}), 
$ M_c(U)$ and  $M_m(U)$. From a single experiment one cannot extract both 
(in principle  this could be achieved with two experiments, 
at markedly different geological sites, 
i.e., with very different values of $n_c$).
 
One has to remind  that $M_m(U)$ has been  estimated by extrapolating 
to the whole mantle rather uncertain data
obtained from the upper mantle.
A reasonable  program is thus: 
i)use available geochemical information to 
fix $M_c(U)$ and ii) use  geo-neutrinos to get information  on 
$M_m(U)$, and thus on $H$.  In this way, from the above equations one gets:
\begin{equation}
\label{heat3}
 H(U) = (h/ n_m) [ N(U)  - (n_c - n_m)M_c(U) ]
\end{equation}

A significant experiment should be capable  of discriminating
among models of heat production.  In the  BSE model one expects 
$ H(U)=8.6$ TW. Predictions for the Uranium contribution are between 
6 and 16 TW, the  lower (upper) value corresponding to the chondritic 
(fully radiogenic) model.  A significant  experiment  should 
 thus provide a measurement with a (1$\sigma$) error of about 2-3 TW. 

By considering the full range of estimated  Uranium mass in the crust as 
a $3\sigma$ interval, one has $M_c(U)= (0.30\pm 0.07)10^{17}$ kg.
From eq.(\ref{heat3}),  which tells how to get $H(U)$ once that $N(U)$
 has been measured with  geoneutrinos, one derives the uncertainty 
on the extracted  value of  $H(U)$ :
\begin{equation}
\label{error}
	\Delta H(U)= (\Delta H_1^2 + \Delta H_2^2)^{1/2} \quad ,
\end{equation}
 where:  
\begin{equation}
\label{error1}
	\Delta H_1   = (h/ n_m) \Delta N(U) \quad ,
\end{equation}
and
\begin{equation}
\label{error2}
\Delta H_2= h \frac{(n_c - n_m)}{n_m} \, \Delta M_c(U) \quad .
\end{equation}

Geoneutrino events are obtained from the counts  number $C$ 
after subtracting reactor $R$ and background $B$ events, so that:
\begin{equation}
\label{delta}
 \Delta N = C^{1/2} = ( N + R + B)^{ 1/2}  \quad .
\end{equation}

As demonstrated by KamLAND, background can be negligible in 
the energy region where most of U-events occur.  
The expectations for  various experiments at different sites, 
collected in table \ref{tab1}, suggest the following comments: 

i)KamLAND can reach  an exposure corresponding to  
$10^{32} p \cdot yr$  (and full efficiency)  in a rather  short  time.  
With this exposure,  unless  the nearby reactors power is reduced,   
the total uncertainty on $H(U)$ would $\Delta H(U)\simeq 8$ TW, dominated 
by fluctuations of reactor events.   A ten times large exposure 
(achievable in several years, with KamLAND, or in a much shorter time 
using a mineral oil detector with  Superkamiokande size) becomes most 
interesting. One can estimate  an accuracy of about 3 TW even with the  
present reactor power.  
This would really provide significant information on the radiogenic 
heat production.  Note however that  some 270 Uranium events will 
have to be extracted from a total of about 2000 counts and that 
in the present calculation accidental background has been neglected.

ii)At Gran Sasso, where the reactor antineutrino flux is much smaller, 
an exposure corresponding to  $10^{32} p \cdot yr$  and full efficiency 
should be  reached  in a reasonable time by Borexino.  The uncertainty 
would be $\Delta H(U) \simeq 5$ TW.
 An experiment with ten times more data could  
reduce the uncertainty to about  2 TW.   

iii)As a limiting case, let us consider and ideal site where crust and 
mantle have the same weight in the determining neutrino events, 
so that $N(U)$ is proportional to the total Uranium mass in the crust 
plus mantle.  The ideal place is thus such that $n_c = n_m$, 
so that uncertainty on $M_c(U)$  do not affect the result, see eq.(\ref{error2}).
Presumably  this means a place in the middle of oceans, far away  
from the continental crust. This place is presumably  also remote 
from nuclear plants, so that the error on N is essentially due to 
statistical fluctuations.   Already with $10^{32} p\cdot yr$ one can reach 
an accuracy of about 2 TW.

In conclusion, there are good prospects for reaching an  accuracy  
$\Delta H(U) =(2-3)$ TW and thus fixing an important missing point  on the sources of  terrestrial heat.

\begin{table}[h]
\footnotesize
\caption{The achievable accuracy $\Delta H$ on the determination of 
$U$ contribution to heat flow.
Calculation are performed for $10^{32} p\cdot yr$, 100\% efficiency 
and a survival probability $P_{ee}=0.55$.
Background counts are set to zero.
In the last two lines, calculations are rescaled for an exposure  
of  $10^{33} p\cdot yr$,
the achievable  accuracy being now $\Delta H(33)$. 
Errors on heat flows are in TW.}
\begin{tabular}{|l|ccc|l|}
\hline
   &\multicolumn{3}{|c|}{Location} & \multicolumn{1}{|c|}{Remarks}\\
   & Kamioka & Gran Sasso & Ideal  &  \\
\hline
$n_c$  	& 	52	&	63  &  18 & \\
$n_m$	&	18	&       18  &  18 &\\
\hline
$R$	&	180 {\footnotesize a)}	& 	35 {\footnotesize b)}  & 0 & {\footnotesize Reactor events:}\\
        &               &              &   & {\footnotesize a)from KamLAND data}\\
        &               &              &   & {\footnotesize b)from Raghavan et al. estimate\cite{Ragh}}\\
$B$	&	0	&	0   &	 0 & {\footnotesize  No background is assumed}\\
$N$(best)&	26.4	&	29.7 & 16.2 & {\footnotesize Geo-events from  best BSE estimate}\\
         &             &            &       &{\footnotesize ($M_c(U)=0.3$ and $M_m(U)=0.6$)}\\
$C=R+B+N$(best)& 206.4 & 64.7 & 16.2 & {\footnotesize Total counts}\\
\hline
$\Delta N= \sqrt{C}$ & 14.4 & 8.04 & 4 &  {\footnotesize Statistical fluctuations only}\\
$ \Delta H_1= (h/n_m) \Delta N$ & 7.6 & 4.2 & 2.1 &  {\footnotesize Error from geo-neutrinos}\\
$\Delta H_2=h (n_c-n_m)/ n_m \, \Delta M_c$ & 1.3 & 1.7 & 0 &  {\footnotesize Error from crust}\\
${\mathbf \Delta H= \sqrt{\Delta H_1^2 +\Delta H_2^2} }$ & {\bf 7.7} & {\bf 4.6} & {\bf 2.1} &  {\footnotesize Total error}\\
\hline
$ \Delta H_1 (33)=(h/n_m) \Delta N $ &  2.4 &  1.3 & 0.7 &  {\footnotesize for $10^{33} p\cdot yr$}\\
${\mathbf \Delta H (33)= \sqrt{\Delta H_1^2 +\Delta H_2^2 }}$ & {\bf 2.7} & {\bf 2.1} & {\bf  0.7} & {\footnotesize for $10^{33} p\cdot yr$}\\
\hline
\end{tabular}
\label{tab1}
\end{table}

\section{Can we learn on neutrinos from geoneutrinos?}

Uncertainties on the 
{ \em individual} Uranium and Thorium  fluxes $\Phi$  are large.  
In practice, one cannot use  $\Phi(U)$ {\em and/or} $\Phi(Th)$  
as additional input for extracting neutrino mass and mixing from  
the lowest energy region
($E_{vis}<E_{geo}=2.6$ MeV)  in a reactor experiment.

However, one can still gain some information on $\theta$ and $\Delta m^2$
 by observing that the {\em ratio} of events from U and Th is well constrained:
\begin{equation}
\label{ratio}
r=N(Th)/N(U)=0.25 \pm 0.05 \quad .
\end{equation}

This  follows from the  fact that the abundance ratio $Th/U$ is well fixed in the solar system, 
being  very similar on meteorites, Venus, Moon as well as on Earth , 
all this information pointing to a common origin of the solar system. 

The  20\%  uncertainty corresponds essentially to uncertainties on the 
$Th/U$ ratio in the various components of the Earth which contribute appreciably 
to neutrino production (continental crust and mantle). This uncertainty accounts 
for  the comparison of different estimates and also of different regions of the Earth interior. 
With respect to the average value $Th/U=3.8$ value, the extrema are $Th/U =2$ for the 
oceanic crust (which is however poor in both $U$ and $Th$ and also it is thin, 
so that it contributes little to neutrino production) and $Th/U \simeq 6$  for the crust estimate 
by one author, most authors giving values in the range 3.8--4.2. 
An estimate $Th/U= 3.8 \pm 0.7$ should be more accurate than 1$\sigma$.

The constraint in eq.(\ref{ratio}) was derived  in \cite{noi2} assuming a uniform  
$Th/U$ distribution inside  the Earth and assuming that  the 
survival probability of geo-neutrinos reaching the detector is the 
same, independently of energy and distance. However, as remarked in 
\cite{Mantovani}, the constraint holds within its error
 even if these simplifying assumptions are relaxed.

Concerning the effect of regional $Th/U$ variations, from
\cite{Chen} it was found  that $r$ is changed by less than 2\% when the
detector is placed at Kamioka, or Gran Sasso, or Tibet (on the top of  a
very thick continental crust) or at the Hawaii (sitting on the middle of a  thin,  $U$-
and $Th$- poor oceanic crust). Coming to the effect of local variations,
by assuming  that within 100 km from the detector the Uranium abundance
is double, $Th/U$=2 , one gets $r=0.22$, whereas if its is halved,
$Th/U$=8, one finds $r=0.28$. Neutrino oscillations clearly do not
affect eq. (7) if the oscillation lengths for both $U$ and $Th$ neutrinos
are both very short or very long in comparison with some typical Earth
dimension. By numerical calculation, one finds that the  effect of 
finite oscillation lengths does not change $r$ by more than 2\% for 
$\Delta m^2 > 1 \cdot 10^{-5}eV^2$. 
In conclusion, all these effects are well within the estimated 20\%
uncertainty on $r$.

The constraint (\ref{ratio}) has been used in \cite{noi2} in order 
to extract information from the full data set of KamLAND.  A slight preference 
for the LMA-I solution with respect to LMA-II has been found, however it is not 
statistically significant. On the other hand,  a significant reduction of the 
mixing parameters space has been found, in spite of the limited statistics. 
This constraint could become really useful when more data are available.

\begin{figure}
\centering
         \mbox{\epsfig{figure=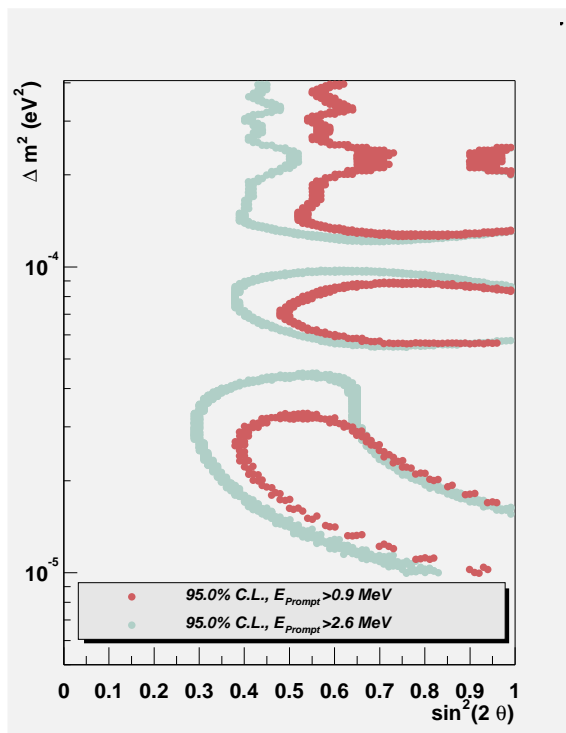,width=7.5cm}}
\caption{The effect of the $Th/U$ constraint on the determination 
of mass and mixing parameters, from ref.$^{10}$}
\label{fig3}
\end{figure}

\section{Concluding remarks: from neutrons to neutrinos}

We all owe very much to Bruno Pontecorvo,  the  man who addressed the detectability 
of neutrinos, discussed  available sources (Sun, reactors and accelerators), 
conceived the Cl-Ar method and invented the beautiful phenomenon of neutrino oscillations. 
There is an additional lesson we can learn from him.  
In 1941, a few years after the celebrated Rome studies on slow neutrons, Bruno was   
a research physicist at Well Surveys Inc.,  and   published a paper entitled 
``Neutron Well Logging - A New Geological Method Based on Nuclear Physics''\cite{Pontecorvo}. 
He had invented the neutron well log, an instrument still used for the 
prospection of water and hydrocarbons, see the page of the   
Society of Professional Well Log Analysts at http://www.spwla.org/. 
It consists of a neutron source and  neutron or gamma detector 
(well shielded from the rays coming directly from the source) to be placed in the well. 
As hydrogen atoms are by far the most effective in the slowing down of neutrons, 
the detected radiation is primarily determined by the hydrogen concentration, 
i.e. water and hydrocarbons.  In this way, the discoveries of the Rome group were  
applied to the study  of  quite different problems. Possibly now we have similar 
opportunities  with neutrinos, due to the important achievements of the last few years:

i)The fate of neutrinos is now essentially understood. We don't  need anymore to rely on standard solar
 model calculations or on the comparison among different experiments. SNO has directly 
observed  the transmutation of solar neutrinos and KamLAND has neatly confirmed the 
result with man made antineutrinos from nuclear reactors.  

ii) KamLAND   has  demonstrated  that it can reach purity levels such that detection 
of geo-neutrinos is feasible, a most impressive experimental result.

All this opens the road for using low energy neutrinos as real  
probes of nature, be it Earth,  Sun,  future and past supernovae 
as well as - may be - the big bang.
Measurement of  the radiogenic  component of terrestrial heat will be the first step. 
It can be achieved with few kiloton (mineral oil) detectors 
in a few years, best if far from nuclear reactors.  

This goal 
can be reached without submerging San Marco inside megaton detectors 
and it will provide  a definite answer to an important and long standing scientific problem.

\section{Acknowledgements}

We are grateful to L. Carmignani, T. Mitsui, T. Lasserre, F. Riccobono, S. Sch\"onert, and A. Suzuki  for useful comments
and suggestions.

\end{document}